\documentclass[letter]{aa}

\usepackage{graphicx}
\usepackage{epsfig}
\usepackage{graphics}
\usepackage{subfigure}
\usepackage{amsmath}
\usepackage{enumerate}
\usepackage{natbib}

\begin{document}

\title{Coronal origin of the polarization of the high-energy emission of Cygnus X-1}

\author{G. E. Romero\inst{1,2,}\thanks{Member of CONICET, Argentina} \and F. L. Vieyro\inst{1,2,}\thanks{Fellow of CONICET, Argentina} \and S. Chaty\inst{3,4}}
  
\institute{Instituto Argentino de Radioastronom\'{\i}a (IAR, CCT La Plata, CONICET), C.C.5, (1984) Villa Elisa, Buenos Aires, Argentina \and Facultad de Ciencias Astron\'omicas y Geof\'{\i}sicas, Universidad Nacional de La Plata, Paseo del Bosque s/n, 1900, La Plata, Argentina \and AIM, UMR-E 9005 CEA/DSM-CNRS-Universit\'e Paris Diderot, Irfu/Service d'Astrophysique, Centre de Saclay,
91191 Gif-sur-Yvette Cedex, France \and Institut Universitaire de France, 103 boulevard Saint-Michel, 75005 Paris, France}

\offprints{F. L. Vieyro \\ \email{fvieyro@iar-conicet.gov.ar}}

\titlerunning{Polarization of Cygnus X-1}

\authorrunning{G.E. Romero et al.}

\abstract
{Cygnus X-1 is the candidate with the highest probability of containing a black hole among the X-ray binary systems in the Galaxy. It is also by far the most often studied of these objects. Recently, the International Gamma-Ray Astrophysics Laboratory Imager onboard then Integral satellite ({\it INTEGRAL}/IBIS) detected strong polarization in the high-energy radiation of this source, between 400 keV and 2 MeV. This radiation has been attributed to a jet launched by the black hole.}
{We consider whether the corona around the black hole might be the site of production of the polarized emission instead of the jet. }
{We studied self-consistently the injection of nonthermal particles in the hot, magnetized plasma around the black hole.}
{We show that both the high-energy spectrum and polarization of Cygnus X-1 in the low-hard state can originate in the corona, without needing to invoke a jet. We estimate the degree of polarization in the intermediate state, where there is no jet, to provide a tool to test our model.}
{Contrary to the commonly accepted view, the jet might not be the source of the MeV polarized tail in the spectrum of Cygnus X-1.}
 
 
\maketitle

\section{Introduction}

Cygnus X-1 is a well-studied X-ray binary. The system is composed of a black-hole candidate of  $14.8 M_{\odot}$ and an early-type O9.7 Iab star of $\sim 20 M_{\odot}$ \citep{orosz2011}. A radio jet was discovered by \citet{stirling2001} and nonthermal high-energy emission was observed by different instruments, from hundreds of keV to GeV, and even perhaps TeV energies \citep{McConnell2000,McConnell2002,albert2007,bodaghee2013,sabatini2010,sabatini2013}.

Recently, \citet{laurent2011} reported the detection and measurement of polarization in the high-energy tail of the emission of Cygnus X-1 detected by the IBIS instrument of the {\it INTEGRAL} satellite. The linear polarization between $400$ keV and $2$ MeV is $67\pm30$ \%. The polarization at lower energies,in contrast, is quite low. This has been interpreted as evidence of a gamma-ray jet in the system \citep{hardcastle2011}. The MeV flux would be, in this interpretation, synchrotron radiation produced by ultra-relativistic primary electrons accelerated in situ close to the base of the jet \citep{zdziarski2012}.

In this work we propose a different scenario for the origin of a polarized MeV tail:  secondary leptonic emission induced by nonthermal particle injection in the hot and magnetized corona around the black hole. Proton acceleration is easier to achieve in such an environment. Then, protons interact with the thermal gas and photons injecting pions and pairs by Bethe-Heitler mechanism. The decay of pions injects muons, which in turn results in more pairs. Absorption of gamma-rays from neutral pion decay is another source of electrons and positrons. As a consequence of all these interactions and decays, a population of energetic secondary pairs appears in the corona. These particles cool mainly by synchrotron emission. The emission becomes partially polarized since part of the magnetic field, which is attached to the accretion disk and twisted around the black hole, is ordered. 

In this letter we present the results of our calculations of nonthermal particle interactions in the corona of Cygnus X-1, and we provide estimates of the secondary leptonic spectra, the MeV emission, and the polarization for both the low hard (LH) and the intermediate soft (IS) states of the source.

\section{Corona of Cygnus X-1}\label{corona}

The existence of a corona of hot (T$\sim 10^9$ K) plasma in Cygnus X-1 was first proposed by \citet{bisnovatyi1977}. If the corona is supported by the magnetic field and the escape of the particles occurs by diffusion instead of advection to the black hole, there is time for thermalization, and ions and electrons share the same temperature, in contrast the advection-dominated cases \citep[ADAFs,][]{narayan1995b}.

The comptonization of disk photons by hot electrons in the corona produces the power-law observed in hard X-rays up to $\sim$ 150 keV \citep{dove1997,poutanen1998}. The existence of a corona is strongly supported by the detection of a Compton reflection feature and the $6.4$ keV Fe K$_\alpha$ line. 
  
The most likely shape of the corona is spherical \citep{dove1997}. In the LH state, the size of the corona lies within $\sim 20-50$ $r_{\rm{g}}$ \citep{poutanen1998}, where $r_{\rm{g}}$ is the gravitational radius. Here we model the coronal region  as a sphere centered on the black hole with a size of $R_{\rm{c}} = 30r_{\rm{g}}$, and a luminosity of 1 \% of the Eddington luminosity, that is, $L_{\rm{c}} = 1.6 \times 10^{37}$ erg s$^{-1}$. 

The magnetic field strength is obtained considering equipartition between the magnetic energy density and the bolometric photon density of the corona \citep[e.g.,][]{bednarek2007,flor01}. This yields $B\sim 5.7\times 10^5$ G.
	
			


			

The hard X-ray emission of the corona is a power-law in photon energy $\epsilon$ with an exponential cut-off at high energies ($	n_{\rm{ph}}(\epsilon) \propto \epsilon^{-\alpha}  e^{-\epsilon/\epsilon_{\rm{c}}}$ erg$^{-1}$ cm$^{-3}$). We adopt $\alpha=1.6$ and $\epsilon_{\rm{c}}=150$ keV, as determined for Cygnus X-1 \citep{poutanen1997}. The photon field of the accretion disk is modeled as a blackbody of temperature $kT_{\rm{d}}=0.1$ keV. 


			

\section{Particle transport}\label{Model}
 
Several works have been devoted to study the effects of nonthermal electron injection in a hot, magnetized corona \citep[e.g.,][]{belmont2008,malzac2009,vurm2009}. Models considering both relativistic electrons and protons have been developed by \citet{flor01} and \citet{vieyro2012}. We adopt here a variation of the latter.

\subsection{Injection} 
		
The mechanism of nonthermal particle injection in black hole coronae is likely to be fast magnetic reconnection: a topological reconfiguration of the magnetic field caused by a change in the connectivity of the field lines. A first-order Fermi mechanism takes place within the reconnection zone, caused by two converging magnetic fluxes of opposite polarity that move toward each other with a velocity $v_{\rm{rec}}$. The resulting injection function of relativistic particles is a power-law $N(E) \propto E^{-\Gamma}$ with an index somewhere in the range $1 \leq \Gamma \leq 3$ \citep{drury2012,bosch-ramon2012}. The best fit of the IBIS data is obtained with $\Gamma = 2.2$, which is consistent with all simulations implemented so far \citep[e.g.,][]{kowal2011}.

The acceleration rate $t^{-1}_{\rm{acc}}=E^{-1}dE/dt$ for a particle of energy $E$ in a magnetic field $B$ is $t^{-1}_{\rm{acc}}=\eta ecB/E$, where $\eta$ is a parameter that characterizes the efficiency of the mechanism in the magnetized plasma. We estimate $\eta \sim 10^{-2}$ (following \citealt{vieyro2012}).

The power available in the system for accelerating particles to relativistic energies can be estimated as in  \citet{delValle2011}, and it yields $\sim 13$ \% $L_{\rm{c}}$. The total power injected into relativistic protons and electrons, $L_{\rm{rel}}$, is assumed to be a fraction of the luminosity of the corona, $L_{\rm{rel}} = q_{\rm{rel}} L_{\rm{c}}$, with the constraint $q_{\rm{rel}} < 0.13$. The way in which energy is divided between hadrons and leptons is unknown. We consider a model where the power injected in protons, $L_{p}$, is 100 times the power in leptons, $L_{e}$. The injection function is both homogeneous and isotropic. The main parameters of the corona model for the LH state of Cygnus X-1 are given in Table \ref{table1}.


%
%
%
%

\begin{table}
    \caption[]{Main parameters of the corona of Cygnus X-1 in the LH state.}
   	\label{table1}
   	\centering
\begin{tabular}{ll}
\hline\hline 
Parameter & Value\\ [0.01cm]
\hline   
$M_{\rm{BH}}$:  black hole mass [$M_{\odot}$]										& $14.8$\tablefootmark{a}	\\
$R_{\rm{c}}$:   corona radius [$r_{\rm{g}}$] 										& $30$ 	                	\\
$r_{\rm{in}}/R_{\rm{c}}$: inner disk/corona ratio               & $0.9$										\\
$\epsilon_{\rm{c}}$:   X-ray spectrum cut-off [keV]							& $150$       					\\
$\alpha$: 			X-ray spectrum power-law index    							& $1.6$									\\
$\eta$: 				acceleration efficiency 												& $10^{-2}$							\\
$B$: 	magnetic field [G]				 																& $5.7 \times 10^5$			\\
$kT$:						disk characteristic temperature [keV] 	  			& $0.1$									\\
\hline 
Free Parameters &  \\
\hline
$n_{i},n{e}$:   plasma density [cm$^{-3}$] 											& $6.2 \times 10^{13}$	\\
$\Gamma$: primary injection index    							                    & $2.2$			  \\
$q$: 						fraction of power injected in relativistic particles 	& $0.05$    	\\
\hline  \\[-0.5cm]
\end{tabular}
\tablefoot{
\tablefoottext{a}{\citet{orosz2011}}}
\end{table}

\subsection{Secondaries}

Primary protons are mainly cooled by photomeson production and, at low energies, by $pp$ interactions. These processes end by injecting secondary leptons and gamma-rays. Secondary pair production is also due to photon-photon annihilation and the Bethe-Heitler process. The most important background photon field for pair creation is the thermal X-ray radiation of the corona. 


\subsection{Radiative losses}

Radiative losses include synchrotron radiation, IC scattering, and relativistic Bremsstrahlung for electrons and muons. Photon production by pair annihilation must also be considered. For protons, the relevant mechanisms are synchrotron radiation, photomeson production, and hadronic inelastic collisions. Non-radiative losses are dominated by diffusion. 

A complete discussion of the cooling times of all these processes can be found in \citet{flor01} and \citet{vieyro2012}. The injection of secondary particles, such as pions and muons, is also discussed in these works.

 The maximum energy for electrons and protons can be inferred using a balance between the acceleration rate and the cooling rate. This yields $E^{e}_{\rm{max}} \sim 10$ GeV for electrons, and $E^{p}_{\rm{max}} \sim 10^{3}$ TeV for protons. Particles of such energies satisfy the Hillas criterion, and can be confined within the corona.

\subsection{Spectral energy distribution}

In this work we aim to determine the equilibrium distribution of secondary pairs. To find it, we solved a system of coupled transport equations in the steady state and assumed spatial homogeneity and isotropy. A detailed discussion of the treatment is presented in \citet{flor01} and \citet{vieyro2012}.

\begin{figure*}
\centering
{\includegraphics[width=0.41\textwidth,keepaspectratio]{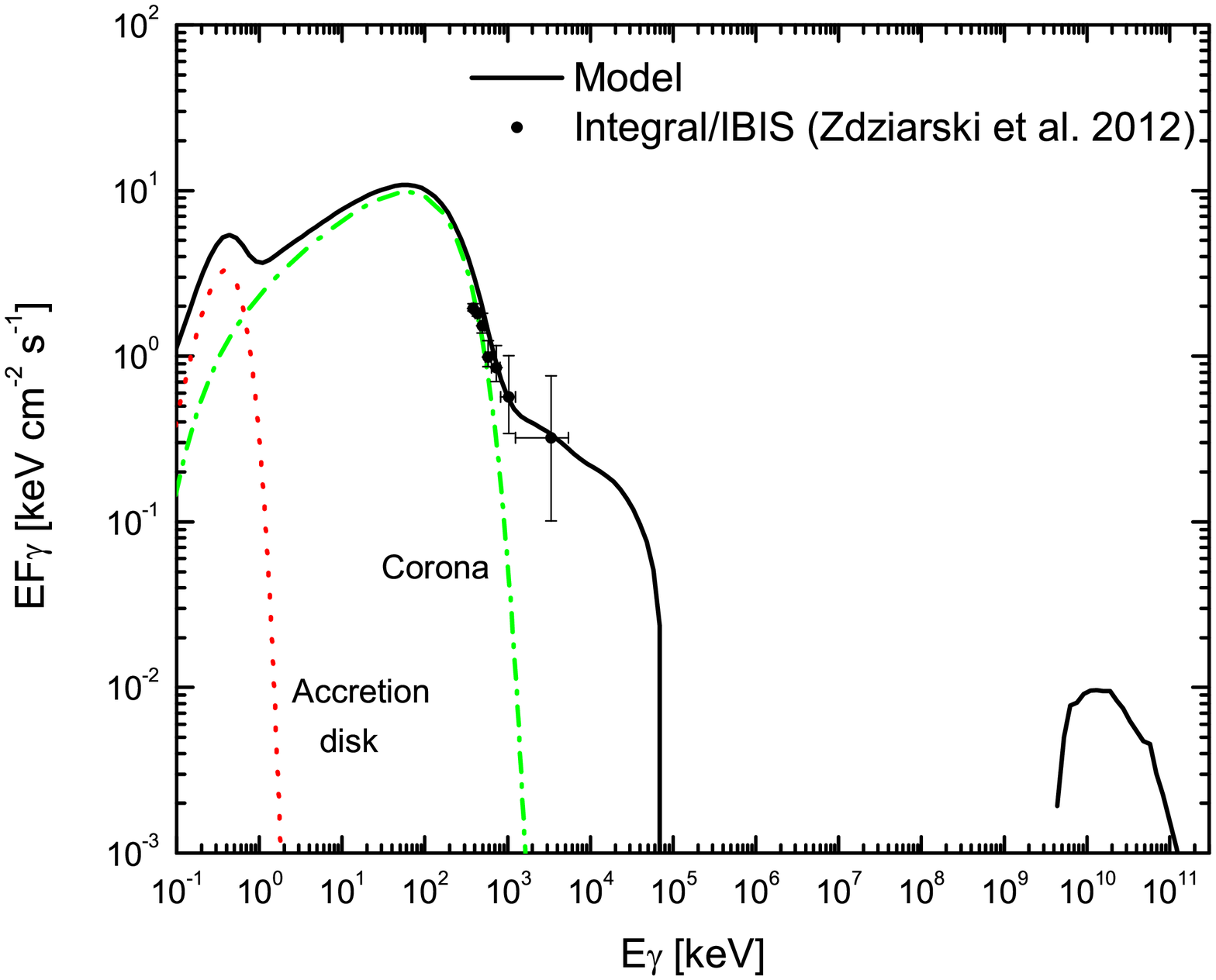}} \hspace{20pt} 
{\includegraphics[width=0.41\textwidth,keepaspectratio]{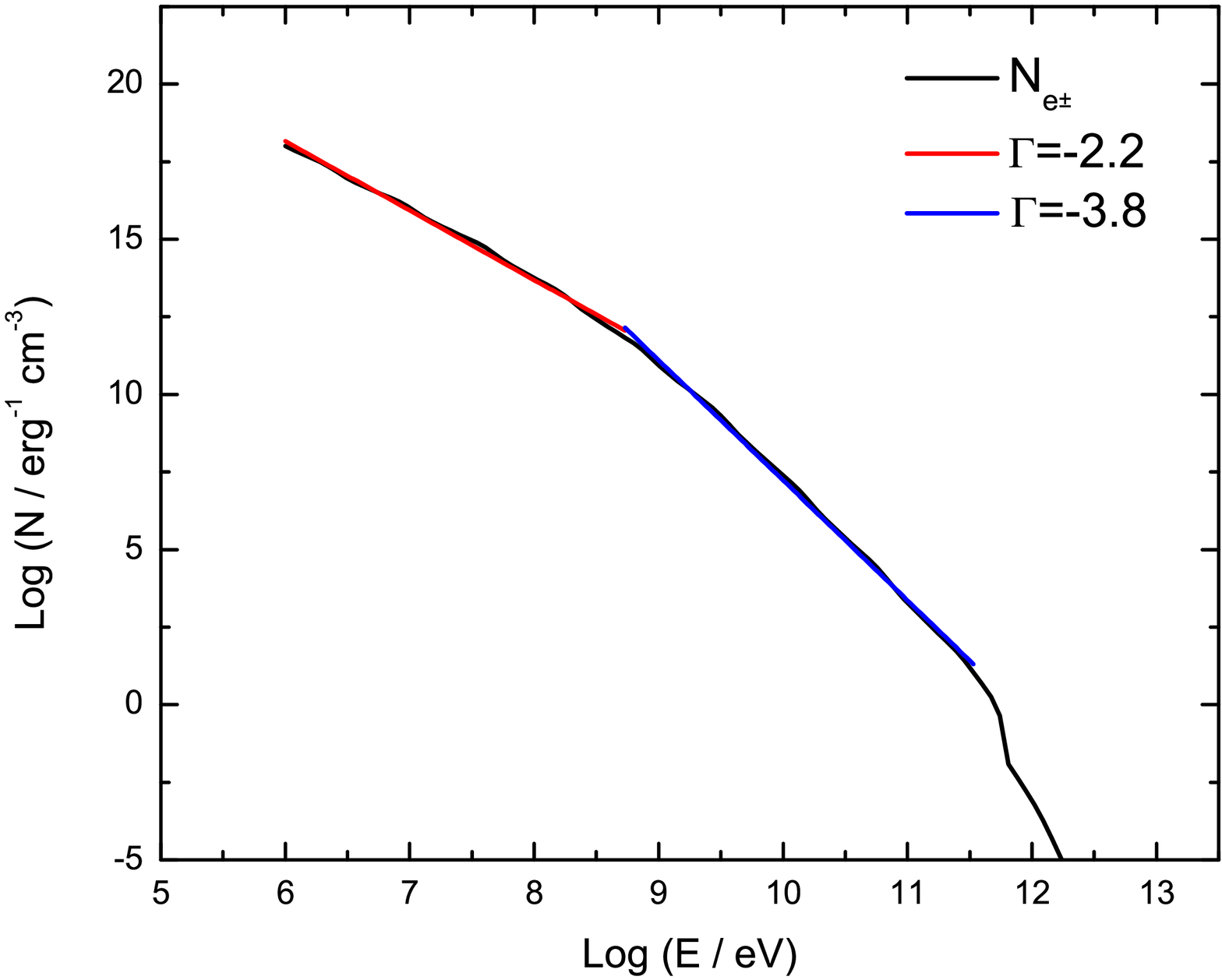}} \hfill 
\caption{Computed SED and IBIS data (left panel), and secondary-pair energy distribution (right panel) for the LH state.}
\label{fig:SED1}
\end{figure*}

The left panel of Fig. \ref{fig:SED1} shows the spectral energy distribution along with the corrected {\it INTEGRAL}/IBIS data \citep{zdziarski2012}\footnote{As noted and corrected by \citet{zdziarski2012}, there is a calibration problem with the original spectrum presented by  \citet{laurent2011}.}. The synchrotron radiation of electron/positron pairs dominates the spectrum for $E_{\gamma} < 100$ MeV. Given the small size of the corona, the synchrotron radiation below $E < 1$ eV is self-absorbed. All radio and infrared emission of the source comes from the jet \citep{stirling2001,zdziarski2012,zdziarski2013,russell2013}.



The synchrotron secondary emission provides an excellent fit of the MeV nonthermal tail of Cygnus X-1. Above 100 MeV almost all emission is absorbed by the coronal X-ray field, with dominance of stellar photons at higher energies ($E > 100$ GeV, see \citealt{vieyro2012}). Radiation in this energy range, as discussed by \citet{zdziarski2013}, is expected be produced in the jet. Hadronic models of jet-wind interaction \citep{romero2003} might be relevant in this energy range as well. 


In the right panel of Fig. \ref{fig:SED1} we present the resulting nonthermal electron/positron distribution in the corona. The spectrum can be fitted with a broken power-law with indices $2.2$ at low energies and $3.8$ at high energies. In contrast to the primary electrons, the most energetic pairs reach energies of $\sim 1$ TeV.

\section{Polarization}\label{Pol}

A relativistic electron/positron population with a power-law energy spectrum $ N(E) \propto E^{-\Gamma} $ will produce a synchrotron flux density $S(\nu) \propto  B_0^{(1+\Gamma)/2} \nu^{(1-\Gamma)/2}$ for an optically thin source with a uniform magnetic field $B_0$ ($E=h\nu$, where $\nu$ is the frequency of the radiation). In this case, the degree of linear polarization will be \citep[e.g.,][]{pacholczyk1970}

\begin{equation}
P_0(\Gamma) = \frac{3\Gamma+ 3}{3\Gamma+ 7}.
\end{equation}

\noindent For an index of $\Gamma=3.8$, as obtained in the previous section for the high-energy secondary pairs, we derive a polarization of $P_0\sim 78.3$ \%. This value is the highest possible, since the turbulent magnetic field will decrease the degree of polarization. Additional changes are introduced if the ordered field is not homogeneous (see \citealt{korchakov1962}). If the random component of the field is $B_{\rm r}$, and the ordered field has components $B_{\perp}$ and $B_{\parallel}$ normal and parallel to the accretion disk, the degree of polarization becomes

\begin{equation}
P_{\rm total}(\Gamma) = \frac{15}{8} \frac{\Gamma+5}{\Gamma+7} P_0\frac{B_0^2}{B_0^2+ B_{\rm r}^2} \frac{\left\langle \Delta B^2 \right\rangle}{B_0^2},
\end{equation}

\noindent with $\left\langle \Delta B^2\right\rangle= \left\langle B_{\perp}^2 \right\rangle-\left\langle B_{\parallel}^2 \right\rangle$ and the average is carried out over the whole range of accessible angles. In the LH state the ejection of a jet requires a minimum inclination angle of the field lines with respect to the axis perpendicular to the disk of $\theta=30$ deg, and the dominant poloidal field is nearly parabolic \citep{romeroVila2013}. Averaging from 30 to 90 deg, we obtain  $\Delta B^2 \sim 0.2 B^2_0 $. 
The observations by \citet{laurent2011} are consistent with a value $P_{\rm total}\sim 50$\%. This would imply $B_{\rm r}\approx 0.73 B_0$, with a total magnetic field $B_0 +  B_{\rm r}\sim 5.7\times 10^5$ G (Sect. \ref{corona}).

\section{Intermediate state}

Cygnus X-1 is most of the time (up to 90\%) in the LH state, but from time to time it enters the so-called intermediate state (IS), when the source transits from LH to high-soft (HS) State.
In this IS, the source exhibits a relatively soft X-ray spectrum ($\alpha \sim2.1-2.3$) and a moderately strong thermal component (see \citealt{malzac2006}).

We applied the model described in the previous sections to the data presented in \citet{malzac2006} (see Fig. 2). We assumed that the luminosity of the corona is 2\% of the Eddington luminosity, which results in $L_{\rm{c}} \sim 3.8 \times 10^{37}$ erg s$^{-1}$. This is in accordance with the count rate of Cygnus X-1 by {\it RXTE} during the period of observation \citep{malzac2006}. The hard X-ray emission of the corona is characterized by a power-law of index $\alpha \sim 2.2$ and an exponential cut-off at $\epsilon_{\rm{c}} = 200$ keV in the IS.

The data from {\it INTEGRAL} do not cover the energy range of the accretion disk emission. Therefore, we adopted $kT_{\rm{max}} = 0.3$ keV, which is an intermediate value between the typical $0.1$ and $0.6$ keV in the LH and HS states, respectively \citep{malzac2006}. The disk inner radius was estimated according to \citet{vila2012}; we considered a luminosity of the disk of $L_{\rm{d}} = 4 \times 10^{37}$ erg s$^{-1}$ \citep{poutanen1998}. The disk/corona ratio was taken as $0.8$, yielding $R_{\rm{c}} \sim 10 r_{\rm{g}}$. 

\begin{figure*}
\centering
{\includegraphics[width=0.41\textwidth,keepaspectratio]{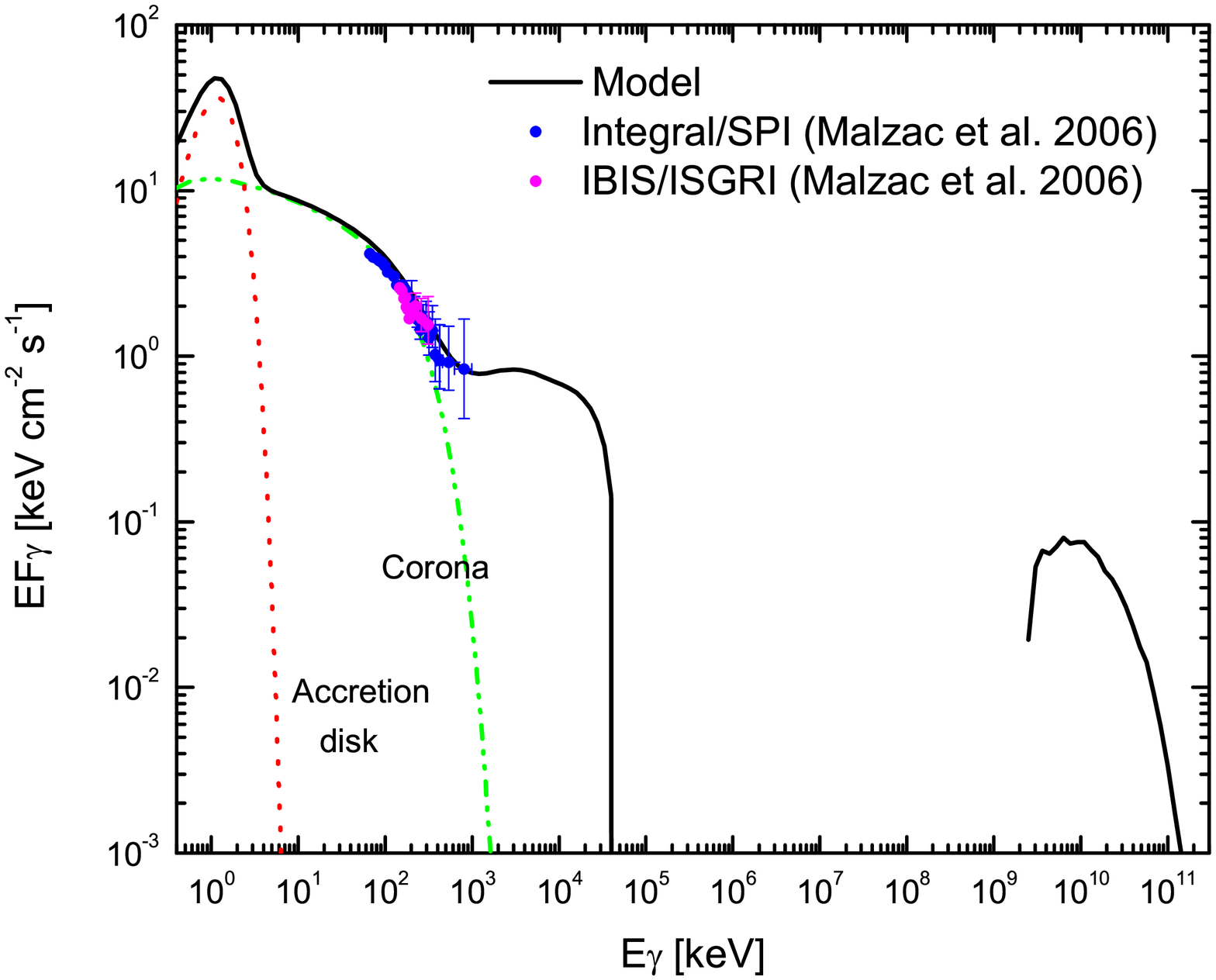}} \hspace{20pt} 
{\includegraphics[width=0.41\textwidth,keepaspectratio]{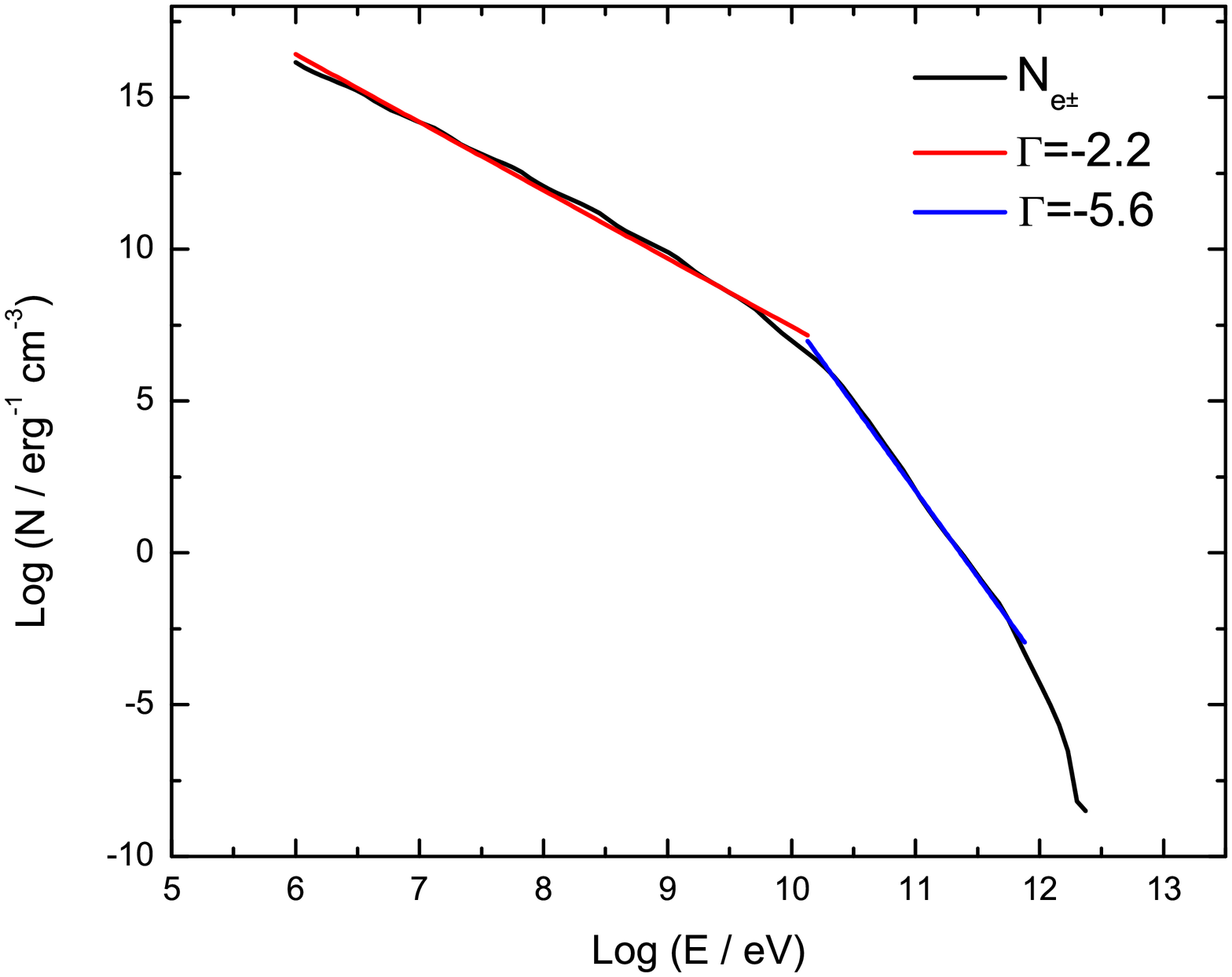}} \hfill 
\caption{Computed SED and SPI+IBIS data (left panel), and secondary-pair energy distribution (right panel) for the IS.}
\label{fig:SED2}
\end{figure*}

%
%
In the right panel of Fig. \ref{fig:SED2} we show the secondary-pair spectrum, which is softer at high energies than the corresponding spectrum in the LH state. In the IS there is no jet, and the accretion disk is much closer to the black hole. This suggests a small inclination angle for the field lines. Adopting a homogeneously ordered field within the Alfv\'en radius, we obtained a linear polarization of $P_{\rm total}=83$ \% $\frac{B_0^2}{B_0^2 + B_{\rm r}^2}$ from the calculated electron/positron spectrum. In this state the expected polarization is $\sim 54$ \%, slightly higher than in the LH state.  In a jet model the polarization in the IS should be zero, since there is no jet. This is a specific prediction of our model.


\section{Conclusions}\label{discussion}

We estimated the polarization of the emission predicted by a hybrid thermal/nonthermal lepto-hadronic corona model for Cygnus X-1. This provides an alternative explanation of the high polarization observed by \citet{laurent2011}. The current data correspond only to the LH state, but our predictions will be testable also with data from the IS state. If polarization is found in the IS, the coronal model will be strongly supported. This can be tested in the future using the polarimetric
capabilities of the soft gamma-ray detector onboard {\it ASTRO-H}.

\section*{Acknowledgments}

This work was supported by the Argentine Agencies CONICET (PIP 0078) and ANPCyT (PICT 2012-00878), as well as by grant AYA2010-21782-C03-01 (Spain). S.C. and G.E.R. acknowledge funding by the Sorbonne Paris Cit\'e (SPC), Scientific Research Project - Argentina 2013. S.C. thanks the Centre National d'Etudes Spatiales (CNES). This work is based on observations obtained with MINE: the Multi-wavelength INTEGRAL NEtwork.

\bibliographystyle{aa}  
\bibliography{myrefs5}   

\end{document}